# High on/off ratio nanosecond laser pulses for a triggered single-photon source

Gang Jin[1, 2], Bei Liu[1, 2], Jun He[1, 2, 3], and Junmin Wang[1, 2, 3, *]

*1. State Key Laboratory of Quantum Optics and Quantum Optics Devices (Shanxi University),*

*2. Institute of Opto-Electronics, Shanxi University,*

*3. Collaborative Innovation Center of Extreme Optics (Shanxi University),*

*No.92 Wu Cheng Road, Tai Yuan 030006, Shan Xi Province, People's Republic of China*

* E-mail: wwjjmm@sxu.edu.cn

An 852nm nanosecond laser pulse chain with a high on/off ratio is generated by chopping a continuous-wave laser beam using a Mach–Zehnder-type electro-optic intensity modulator (MZ-EOIM). The detailed dependence of the MZ-EOIM's on/off ratio on various parameters is characterized. By optimizing the incident beam polarization and stabilizing the MZ-EOIM temperature, a static on/off ratio of 12600:1 is achieved. The dynamic on/off ratios versus the pulse repetition rate and the pulse duty cycle are measured and discussed. The high-on/off-ratio nanosecond pulsed laser system was used in a triggered single-photon source based on a trapped single cesium atom, which reveals clear antibunching.

Integrated optical devices play important roles in high-speed fiber telecom and laser technology owing to advances in flexibility, small size, high sensitivity, and other characteristics. Lithium niobate ($LiNbO_3$) optical waveguide devices can provide ultrafast modulation of a laser's amplitude, phase, and frequency.[1,2] For a $LiNbO_3$ waveguide Mach–Zehnder-type electro-optic intensity modulator (MZ-EOIM), square-wave electrical pulses applied to the modulation port can slice a continuous-wave laser beam into a rectangular wave with a typical pulse duration on the order of nanoseconds or even less, which could serve as an excitation laser to implement a triggered single-photon source.[3–5] To achieve a nearly deterministic triggered single-photon source, it is necessary to realize excitation laser pulses with a high on/off ratio. However, because of the limited fabrication technology[6] used to manufacture MZ-EOIMs, polarization fluctuations of the incident laser and thermal drift may cause significant fluctuation of the output pulses at minimum intensity, which is defined as the OFF state, where the maximum transmission is defined as the ON state.

As shown in Fig. 1, a single cesium (Cs) atom is trapped in a microscopic optical tweezer[7] in the ground state $6S_{1/2}$, $|F_g = 4, mF=+4>$. When a resonant $\sigma+$-polarized $\pi$ pulse (pulse duration $\tau_e$ = 5 ns) is applied, the atom can be excited to $6P_{3/2}$ $|Fe=5, mF=+5>$. The spontaneously emitted photons can be collected by a high-numerical-aperture lens assembly. The laser intensity in the OFF state should be closer to zero to reduce the number of multiphoton events in the single-photon source. Therefore, increasing the ratio of the maximum and minimum powers (on/off ratio) of the output pulsed light is particularly important.

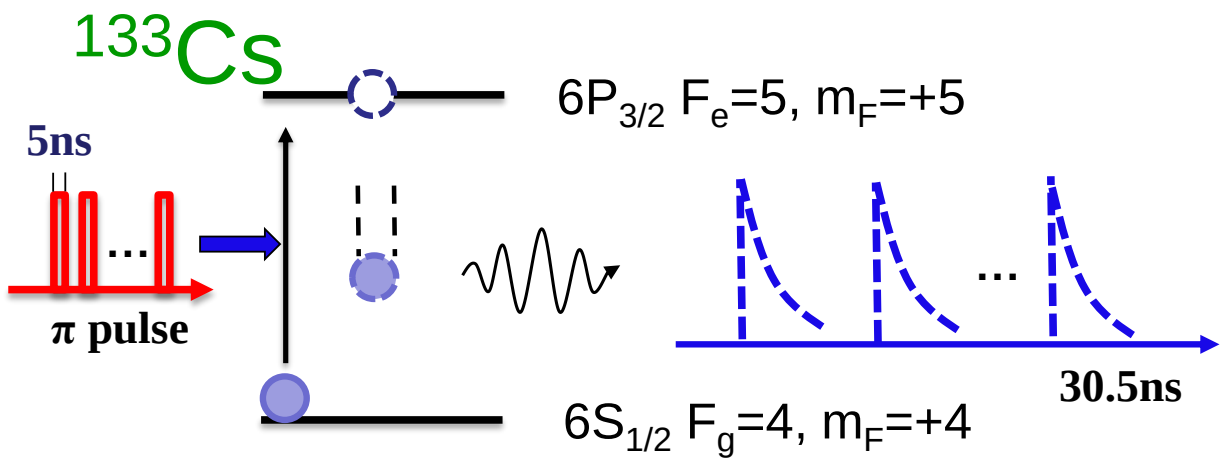

**Fig. 1.** Schematic diagram of single-photon source based on periodic excitation of a trapped single cesium atom.

The on/off ratio of a commercial MZ-EOIM at a wavelength of 850 nm is generally not so good. Two problems need to be addressed regarding the pulse train generated by MZ-EOIM: reducing the power in the OFF state as much as possible and improving power stability. The on/off ratio and power stability of the output pulsed light depend on the symmetry of the two arms of the MZ interferometer, the temperature, the polarization of the input laser, and the external stress.[6)] Various approaches may be used to improve the stability and increase the on/off ratio of an MZ-EOIM. Optimizing the structure of the modulator design[8,9)] can decrease the insert losses and improve the symmetry of the MZ-EOIM. However, this approach requires high technical accuracy. Dingjan et al.,[5)] Snoddy et al.,[10)] and Bui et al.[11)] implemented a feedback loop in real time to recalibrate DC operating point, but this approach would produce additional unwanted intensity fluctuation. The use of cascaded MZ-EOIMs[2,12)] decreases the available output power to achieve a high on/off ratio. Moreover, it is difficult to maintain synchronization in time when the electric signal is applied to two modulators, which is necessary for use in high-frequency signal applications. In this letter, by finely matching the polarization of the input light with a single-mode polarization-maintaining (PM) fiber and precisely stabilizing the temperature, the modulator is made insensitive to temperature[13)] and polarization[14)] fluctuation. Importantly, the output laser's on/off ratio can also be increased.

$LiNbO_3$ is used for electrical control of the optical phase in each arm of the MZ-EOIM owing to its linear electro-optic effect.[15)] According to the principle of the MZ interferometer, when the two branches are combined at a Y junction, the laser intensity can be changed by varying the applied voltage. As shown in Fig. 2(a), the incident light wave $E_0$ is divided into two parts according to the ratio $a^2/(1-a^2)$ at the first Y junction beam splitter, and the phases of the two expected waves propagated in each arm were changed by $\Delta\phi_1$ and $\Delta\phi_2$ by the applied electric field. In addition, the insertion losses differ according to the transmissivities $T_1$ and $T_2$, respectively, of the two arms. When two beams are combined at the second Y junction according to the intensity ratio $b_2/(1-b_2)$, the output E can be written as

$$E = bT_1aE_0\exp(-i\Delta\phi_1) + \sqrt{1-b^2}T_2\sqrt{1-a^2}E_0\exp(-i\Delta\phi_2) \quad \textbf{(1)}$$

In the ideal case, the symmetric arms of the MZ-EOIM can be described by $a^2 = 1/2,\ b^2 = 1/2,\ T_1 = T_2$ . If the phase difference between the two arms is $\Delta\phi = \Delta\phi_2 - \Delta\phi_1$, the output laser intensity is given by

$$I \propto |\mathrm{E}|^2 = \frac{1}{2}(\mathrm{T}_1 E_0)^2 [1 + \cos(\Delta\phi)] \quad \textbf{(2)}$$

In fact, the degradation of the on/off ratio is caused by an imbalance of the two arms in the MZ-EOIM. Here we consider only the effect of the splitting ratio on the results. Assuming $b^2 = 1/2$ and $T_1 = T_2 = 1$, the output light intensity in Eq. (1) can be rewritten as

$$I \propto |E|^2 = E_0{}^2 \left[1 + 2a\sqrt{1-a^2}\cos(\Delta\phi)\right]/2 \quad \textbf{(3)}$$

According to Eq. (3), the minimum intensity is no longer zero when $\Delta\phi = \pi$; thus, the on/off ratio R has a finite value:

$$R = \frac{P_{\max}}{P_{\min}} = \frac{1 + 2a\sqrt{1-a^2}}{1 - 2a\sqrt{1-a^2}} \quad \textbf{(4)}$$

The dependence of the theoretical on/off ratio on the splitting ratio is shown in Fig. 2(b). When the splitting ratio is 50/50, the on/off ratio should be infinite in theory, and it will sharply fall from infinity to below 2000 when the splitting ratio is worse than 52/48. Considering various losses and the splitting ratio between the two arms, the on/off ratio will actually be no more than 1000. Thus, an efficient way to optimize the on/off ratio is necessary.

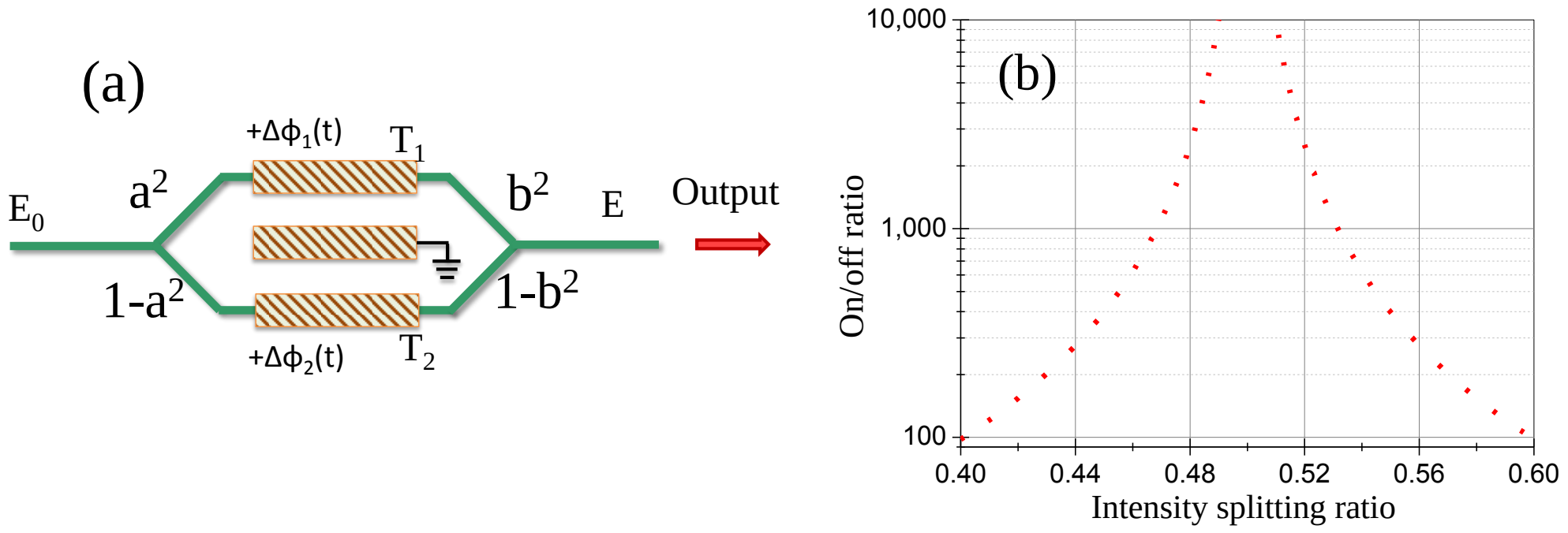

**Fig. 2.** (a) Schematic diagram of MZ-EOIM. (b) Simulated on/off ratio versus intensity splitting ratio of MZ-EOIM.

An extended-cavity diode laser (ECDL; Toptica DL100) provides an 852 nm output with a maximum power of 80mW. The ECDL's frequency is locked to the $6S_{1/2}$ (Fg = 4)→$6P_{3/2}$ (Fe = 5) closed transition by using the polarization spectra[16)] via a proportional-integral amplifier and the piezoelectric transducer driver of the ECDL. The main output beam passes through the MZ-EOIM to generate high-on/off-ratio laser pulses for exciting single Cs atom trapped in our microscopic optical tweezer.[7] Before the incident beam is fed into the MZ-EOIM via the input PM fiber, it is linearly polarized with a high extinction ratio (>40 dB). The incident beam's polarization is adjusted to match the input PM fiber.

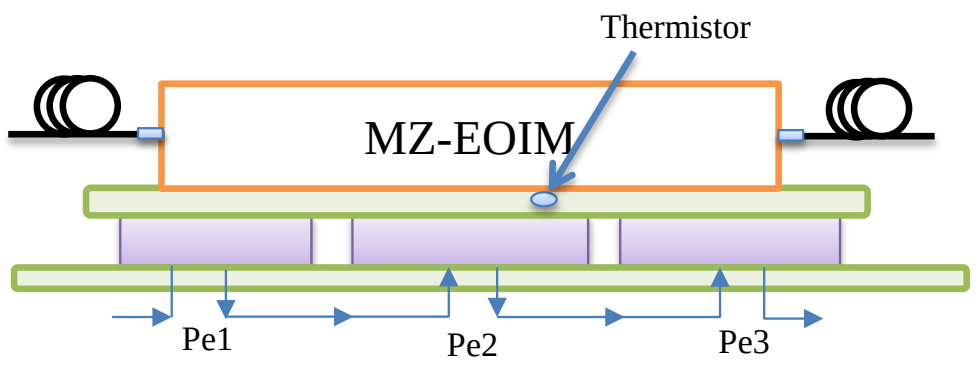

**Fig. 3.** Diagrammatic drawing of temperature-stabilized system. Pe1, Pe2, and Pe3 in series are the Peltier modules, which are employed to heat or cool the EOIM. A thermistor connected to the temperature controller (not shown in this figure) is used to monitor and feedback the MZ-EOIM's temperature value.

$LiNbO_3$ waveguides with various types of crystal cut (X-cut, Y-cut, and Z-cut) are generally fabricated in wafers.[17] The Z-cut mode ridge waveguides increase the overlap between the applied electric field and the laser mode, which provides a lower half-wave voltage but introduces pyroelectric and thermo-optic effects.[15,18] Because the electrodes are located directly above the waveguide, the MZ-EOIM working conditions are strongly affected by temperature. Consequently, the DC bias voltage of the Z-cut MZ-EOIM used in our experiment drifts with the ambient temperature, which is problematic for generation of high-on/off-ratio laser pulses.

To suppress thermal drift, a temperature-stabilized device is designed, as shown in Fig. 3. The MZ-EOIM package is placed on a thin copper plate with thermal grease to improve heat conduction. A 10 kΩ calibrated thermistor is placed inside the copper plate to monitor the MZ-EOIM's temperature. Three parallel Peltier modules (thermal electric coolers) are installed under the copper plate and connected to a temperature controller (Thorlabs TED-200C) to stabilize the temperature. The typical temperature instability is less than 5mK in 24 h. The temperature-controlled MZ-EOIM is covered with an organic glass box to isolate it from air flow and weaken temperature fluctuation.

As described by Eq. (3), we can trace the intensity at the outlet of the MZ-EOIM according to the applied voltage, as shown in Fig. 4(a). The three curves represent different temperatures (296.4, 298.3, and 299.8 K). The maximum output power in the ON state is 7.5mW. However, the minimum output power in the OFF state varies with the DC bias voltage, the incident polarization, and the MZ-EOIM's temperature. After the incident polarization is matched, the power transmission curves at different temperatures will shift in a specific direction. In principle, if there is no applied electronic signal (0 V), the zero relative phase shift should indicate that the transmitted power reaches the maximum value. However, the result in Fig. 4(a) indicates that the temperature difference is less than 3.5K without a DC bias supply (the vertical solid line), but the output power varies greatly, corresponding to a variation in the transmissivity from 99.6 to 45.8%, where the phase shift is caused by a temperature difference due to the imbalance of the two arms, pyroelectric effects, and the temperature gradient between the two arms.[13)]

The temperature-dependent phase shift is shown in Fig. 4(b), the temperature-induced phase shifts are about π/5.68 and π/5.58K at DC bias voltages of 0 and 3V, respectively. This result may be viewed as a consequence of the different arm lengths in the MZ-EOIM or the different thermal expansion induced by the temperature gradient. The temperature-dependent extraordinary index of refraction in $LiNbO_3$ is described by the Sellmeier equation:[19)]

$$n^2 = 5.35583 + 4.629\times10^{-7} f + \frac{0.100473 + 3.862\times10^{-8} f}{\lambda^2 - (0.20692 - 0.89\times10^{-8} f)^2} + \frac{100 + 2.652\times10^{-5} f}{\lambda^2 - 11.34927^2} - 1.5334\times10^{-2}\lambda \qquad \textbf{(5)}$$

where the temperature parameter is $f = (T - T_0)(T + T_0 + 2\times273.15)$, $T_0$ = 297.65 K, λ = 0.852 μm, Further, the phase change $\Delta\phi$ of the MZ-EOIM is generated by the voltage V applied to the electrodes with an interval of d:

$$\Delta\phi = k(\Delta n\cdot L_{DC} + n\cdot\Delta L) = (n^3\gamma_{eff} V L_{DC}\lambda\pi / d + n\, k\Delta L) + \phi_0 \qquad \textbf{(6)}$$

where k = 2π/λ is the wavenumber of the incident beam, the change in refractive index $\Delta n = n^3 E\gamma_{\mathrm{eff}}/2$, the electric field strength $E=V/d$, the effective electro-optic coefficient $\gamma_{\mathrm{eff}}$ = 30.8 pm/V, $L_{\mathrm{DC}}$ is the DC electrode length, and $\Delta L$ and $\phi_0$ represent the length difference and initial phase difference between the two arms of the waveguide, respectively. The open circles and squares in Fig. 4(b) represent the results at DC voltages of 0 and 3V, respectively. The solid lines in Fig. 4(b) represent fittings with Eqs. (3) and (6) and show good agreement with the measured data at room temperature, which is below 300 K. The somewhat large deviation shown at temperatures above 300K may be caused by the temperature gradient between the sensor and the MZ-EOIM waveguide. The inset in Fig. 4(b) shows a double local minimum of the output power, so we fix the lowest local minimum as the OFF state to achieve a high on/off ratio. In fact, the temperature-induced phase shift includes the effects of the asymmetric arms and the pyroelectric field,[10,20] specially for different electrode lengths.[21]

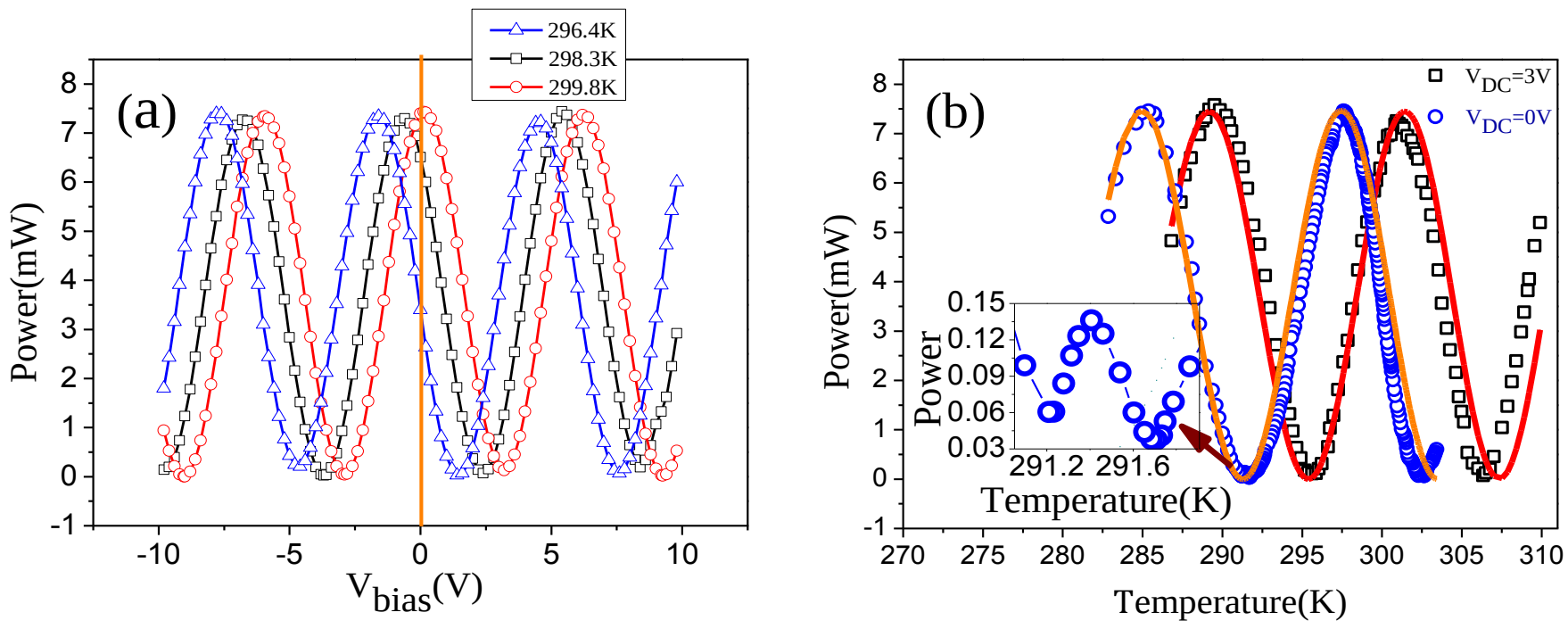

**Fig. 4.** (a) Dependence of the output laser power on the DC basis voltage. Blue triangles, black squares, and red circles represent the output power versus the DC bias at temperatures of 296.4, 298.3, and 299.8 K, respectively. (b) Output laser power of the MZ-EOIM at DC bias voltages of 3 and 0V. Inset shows output power versus temperature.

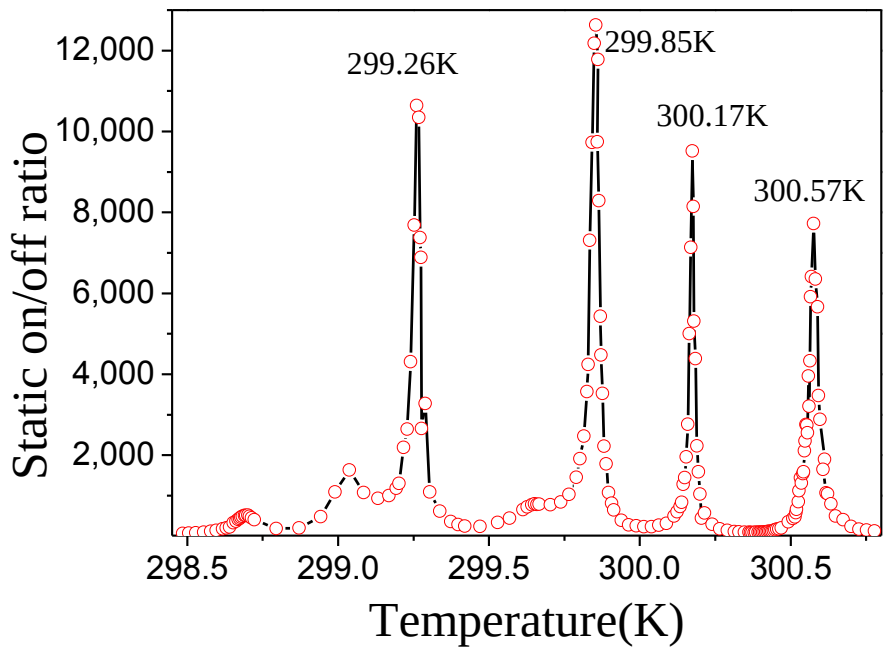

**Fig. 5.** Measured static on/off ratio of MZ-EOIM at different temperatures.

After the incident light polarization is carefully matched with the PM fiber and the preset value of the temperature controller is slowly changed, the minimum power at a specific bias voltage can be determined. If a half-wave voltage is added on the modulation port of the MZ-EOIM, the output changes to the ON state. The measured static on/off ratios are shown in Fig. 5. The measured maximum on/off ratio is as high as 12600: 1, representing an increase by 21 dB compared with a typical value of 100: 1. Every peak in Fig. 5 has a sharp slope, and the profile is similar to the predicted curve profile in Fig. 2(b), indicating that the two arms tend to be perfectly symmetric.

All the measured pulse chains were delivered into single-photon-counting modules (SPCM-AQRH-15-FC) by a multimode fiber after attenuation of the output of the temperature-controlled MZ-EOIM, they were recorded by a P7888 card (FAST ComTec). The acquisition card is triggered by a synchronized electrical pulse to start. Figure 6 shows the dynamic on/off ratio, which is the ratio of the average photon counts per second in the ON state and in the OFF state. At repetition rates of a few kilohertz, the on/off ratio remains above 5000, but it gradually drops to several hundred as the repetition rate reaches the megahertz level [Fig. 6(a)]. The same behavior appears when we increase the pulse duty cycle at a 1MHz repetition rate [Fig. 6(b)], which is caused mainly by negative overshoot of the electronic signal or the charge and discharge between electrodes. After the settling time of the OFF state, the residual photon counts are close to the static conditions.

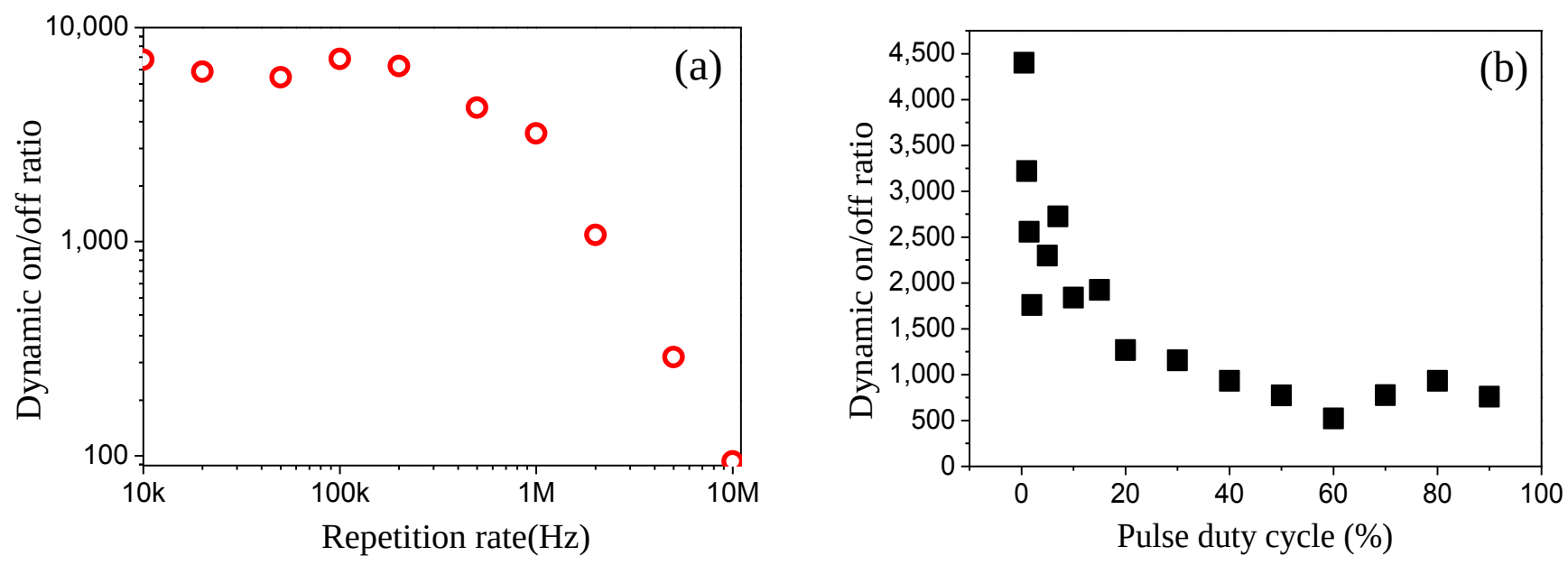

**Fig. 6.** (a) Dynamic on/off ratio at 5 ns pulse duration versus laser pulse repetition rate. (b) Dynamic on/off ratio versus laser pulse duty cycle atrepetition rate of 1 MHz.

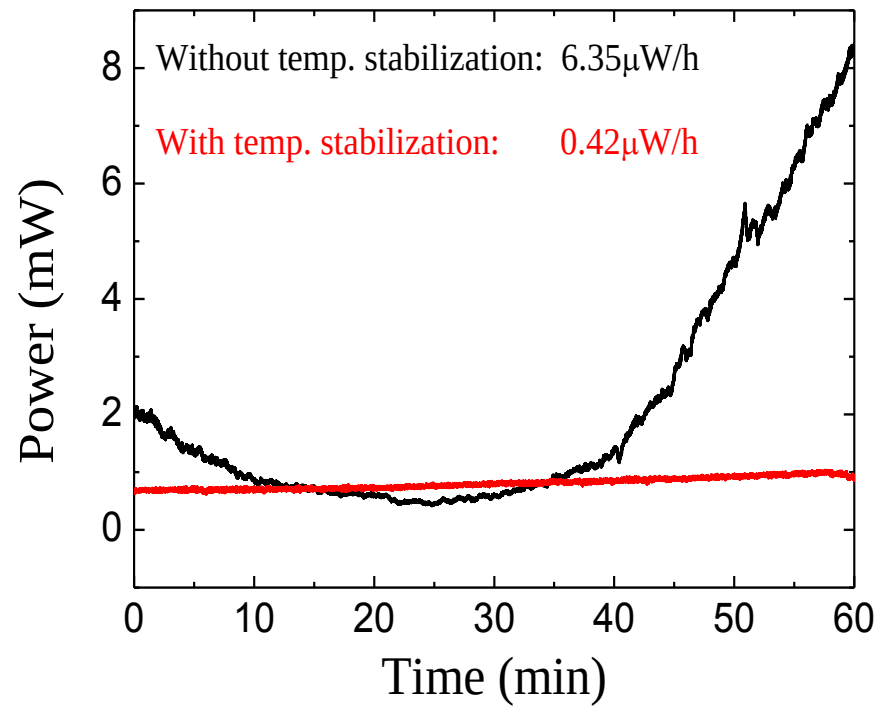

**Fig. 7.** Power stability in OFF state for 1 h with (red solid line) and without temperature stabilization (black solid line).

Because the temperature-dependent refractive index[19,21] causes the phase difference between the two arms to fluctuate with temperature, the DC bias voltage in the OFF state will change. Furthermore, the thermal effect will also deform the waveguide's geometric structure, causing mismatching of the light mode inside the waveguide and affecting the waveguide transmissivity of the arms ($T_1$ and $T_2$). Consequently, the beam splitting ratio at the Y junction deviates from 50/50. The temperature stabilization system can accurately match the temperature to the optimum working point, at which the coefficient is balanced between the two branches; thus, the on/off ratio increases remarkably. The disadvantage is that the optimum temperature should be calibrated manually to the best result because of the long distance to the waveguide chip and Peltier modules, which can be optimized by an intergrade process.

In this method, because the peak intensity fluctuation percentage is much smaller than the OFF state, the power stability in the OFF state reflects the stability of the on/off ratio. As shown in Fig. 7, not only does the MZ-EOIM reach a lower power in the OFF state, but also the power drift can be remark-ably suppressed. In the OFF state, the power drifts of the MZ-EOIM with and without temperature stabilization are 0.42 and 6.35μW/h, respectively. The MZ-EOIM waveguide reaches a steady thermal equilibrium, resulting in a stable output.

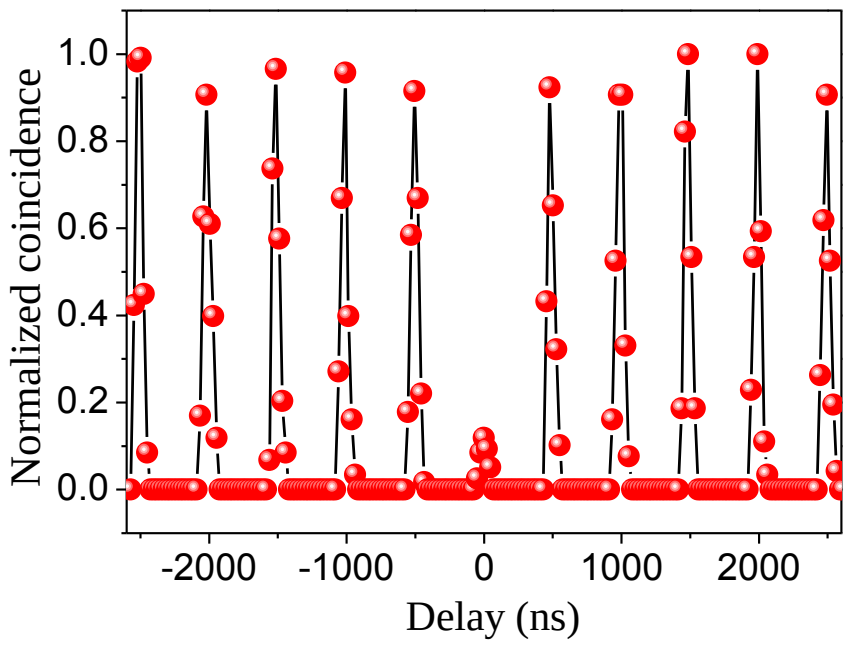

**Fig. 8.** Antibunching of a triggered single-photon source based on a trapped single cesium atom.

To prove the practicability of the high-on/off-ratio pulsed laser, we demonstrate the second-order intensity correlations of fluorescence spontaneously emitted from a trapped single Cs atom excited by a 5 ns pulse chain with a repetition rate of 2MHz. Figure 8 shows the results, which are measured by a Hanbury Brown and Twiss setup[22) without any background subtraction, the coincidence close to zero around zero delay is the signature of a triggered single-photon source.

In conclusion, we provided a practical way to improve two intrinsic defects of commercial MZ-EOIMs, the limited on/off ratio and bias shift. We found that the MZ-EOIM's on/off ratio is sensitive to the thermal shift caused by heat exchange from the thermal bath in the laboratory environment and partial absorption of continuous-wave laser radiation and electronic signals. Temperature stabilization is implemented to reduce heat accumulation. The result indicates that temperature fluctuations seriously affect the symmetric performance of the modulator and further degrade the on/off ratio. In spite of these problems, by choosing the proper value by adjusting the temperature, the symmetry of the MZ-EOIM can be optimized to a near perfect point, and the static on/off ratio can be increased by at least 21 dB. The OFF-state power drift is reduced to 0.42μW/h.

Moreover, this scheme may also be combined with traditional methods of achieving a higher on/off ratio, such as the use of cascaded MZ-EOIMs[2,12] and servo feedback control.[5,10,11] Although the output power is low when cascaded temperature-stabilized MZ-EOIMs are used, if the acquired weak pulsed laser can be applied to an injection-locked slave diode laser with a large detuning, a higher-power laser pulse can be obtained by inserting an etalon to filter unwanted optical wavelengths versus the slave laser free running. This system provides an elegant method of providing excitation laser pulses for a triggered single-photon source based on single atom manipulation,[7] as well as high signal-to-noise time-bin entanglement.[23] Single photons obtained directly from single atoms may more easily enhance the atom–photon interaction in an atomic ensemble and act as a quantum memory owing to the narrow bandwidths (about 5 MHz).[24] The proposed technique is suitable for general asymmetric MZ-EOIMs, and the analysis and discussion may provide some references for MZ-EOIM manufacturers to accelerate the development of quantum information processing.

**Acknowledgements**

This project is supported by the National Major Scientific Research Program of China (Grant No. 2012CB921601) and the National Natural Science Foundation of China (Grant Nos. 61475091, 11274213, 61205215, and 61227902).